\documentclass[twocolumn,english,showpacs,superscriptaddress,prm]{revtex4-1}
\usepackage[colorlinks=true,citecolor=blue,linkcolor=blue,urlcolor=blue]{hyperref}
\usepackage[latin9]{inputenc}
\usepackage{color}
\usepackage{float}
\usepackage{textcomp}
\usepackage{amstext}
\usepackage{graphicx}
\usepackage{gensymb}
\makeatletter
\usepackage{graphics}\usepackage{subfigure}\usepackage{longtable}\usepackage{pstricks}\usepackage{dcolumn}\usepackage{bm}
\usepackage{amsmath}
\begin{document}
\title{Investigation of hydrogen incorporations in bulk infinite-layer nickelates}
\author{P.~Puphal}
\email{puphal@fkf.mpg.de}
\affiliation{Max-Planck-Institute for Solid State Research, Heisenbergstra{\ss}e 1, 70569 Stuttgart, Germany}
\author{V.~Pomjakushin}
\affiliation{Laboratory for Neutron Scattering and Imaging (LNS), Paul Scherrer Institute (PSI), Forschungsstrasse 111, CH-5232 Villigen, Switzerland}
\author{R.~A.~Ortiz}
\affiliation{Max-Planck-Institute for Solid State Research, Heisenbergstra{\ss}e 1, 70569 Stuttgart, Germany}
\author{S.~Hammoud}
\affiliation{Max-Planck-Institute for Intelligent Systems, Heisenbergstra{\ss}e 3, 70569 Stuttgart, Germany}
\author{M.~Isobe}
\affiliation{Max-Planck-Institute for Solid State Research, Heisenbergstra{\ss}e 1, 70569 Stuttgart, Germany}
\author{B.~Keimer}
\affiliation{Max-Planck-Institute for Solid State Research, Heisenbergstra{\ss}e 1, 70569 Stuttgart, Germany}
\author{M.~Hepting}
\affiliation{Max-Planck-Institute for Solid State Research, Heisenbergstra{\ss}e 1, 70569 Stuttgart, Germany}

\date{\today}

\begin{abstract}
Infinite-layer (IL) nickelates are an emerging class of superconductors, where the Ni$^{1+}$ valence state in a square planar NiO$_2$ coordination can only be reached via topotactic reduction of the perovskite phase. However, this topotactic soft chemistry with hydrogenous reagents is still at a stage of rapid development, and there is a number of open issues, especially considering the possibility of hydrogen incorporation. Here we study the time dependence of the topotactic transformation of LaNiO$_3$ to LaNiO$_2$ for powder samples with x-ray diffraction and gas extraction techniques. While the hydrogen content of the powder increases with time, neutron diffraction shows no negative scattering of hydrogen in the LaNiO$_2$ crystal lattice. The extra hydrogen appears to be confined to grain boundaries or secondary-phase precipitates. The average crystal structure, and possibly also the physical properties, of the primary LaNiO$_2$ phase are therefore not noticeably affected by hydrogen residues created by the topotactic transformation.

\end{abstract}

\maketitle

\section{Introduction}

Superconductivity exists in various hydrogen containing compounds, highlighted by the recent discoveries of critical temperatures $T_c$ as high as room temperature for hydride compounds formed under extreme pressures \cite{Shamp2017,Snider2020}. Metal hydroxide-intercalated iron chalcogenides, such as (Li$_{0.8}$Fe$_{0.2}$)OHFeSe \cite{Lu2014}, show coexistence of antiferromagnetic order and superconductivity, a feature known from some high-$T_c$ superconductors \cite{Kahsay2016}. Electron-doped 1111 iron pnictides as $R$FeAsO$_{1-x}$H$_x$ \cite{Hanna2011,Hiraishi2014}, CeFeAsO$_{1-x}$H$_x$ \cite{Matsuishi2012}, and the pnictogen-free LaFeSiH \cite{Bernardini2018} are a class, where the introduction of charge carriers via doping with hydrogen drives the system from an antiferromagnetically ordered state towards superconductivity. In layered  sodium  cobalt oxyhydrate,  Na$_x$CoO$_2\cdot$H$_2$O \cite{Lynn2003}, superconductivity with a similar hole/electron-doping behavior by chemical substitution is observed as in the cuprate high-$T_c$ superconductors \cite{Keimer2015}. 

For IL nickelates, a close relation and possible analogy to cuprate superconductors was suggested already in 1999 \cite{Anisimov1999} and since the first discovery of superconductivity in the IL nickelate (Nd,Sr)NiO$_2$ \cite{Li2019}, the observation of superconductivity has been confirmed \cite{Zeng2020,Li2021G,Gao2021C} and extended to (Pr,Sr)NiO$_2$ \cite{Osada2020}, (La,Sr)NiO$_2$ \cite{osada2021}, (La,Ca)NiO$_2$ \cite{zeng2021}, and Nd$_6$Ni$_5$O$_{12}$ \cite{Pan2021}.
Furthermore, a recent work reported superconductivity not only for films grown on SrTiO$_3$ substrates, but also on LSAT \cite{Ren2021}, which provides enough evidence to consider thin film nickelates as a novel class of superconductors.
The possible presence of topotactic hydrogen in IL nickelates, which depends on the rare-earth ion and/or epitaxial strain \cite{Bernardini2021,Alvarez2021}, was proposed in theoretical studies \cite{Si2020,Malyi2021}, and might have substantial influence on the electronic and magnetic properties of the IL nickelates, as LaNiO$_2$H would realize a two-orbital Mott insulator \cite{Si2020}. A hint towards the possibility of hydrogen incorporation was provided by an early study of topotactically reduced NdNiO$_{3}$ films, which showed an oxyhydride NdNiO$_{3-x}$H$_y$ phase with a defect-fluorite structure in the surface region \cite{Onozuka2016}. Furthermore, topotactic hydrogen can be found in SrTiO$_3$ thin films \cite{Kutsuzawa2018}, \textit{i.e.} the material that is commonly used as substrate and capping layer for IL nickelate films, which again provides a possible route for inclusion of hydrogen in infinite-layer nickelates thin films.
While superconductivity has remained elusive in IL nickelates in bulk form, with studies on powder samples reporting insulating behavior \cite{Wang2020,Li2020}, a first step towards superconductivity was taken by our recent investigation of (La,Ca)NiO$_2$ single crystals \cite{Puphal2021}, where metallicity was observed. 

In this work, we address the issue of possible hydrogen incorporations in IL nickelates by examining bulk LaNiO$_{3-x}$H$_y$ powder samples with a combination of x-ray and neutron diffraction studies, gas extraction, as well as complementary high pressure synthesis attempts of LaNiO$_{2}$H.

\section{Methods}

LaNiO$_{3}$ powder samples with grain sizes of 0.5 $\mu$m were synthesized via the citrate-nitrate method as described in Ref. \cite{Ortiz2021}. The method is optimal as the relatively small grains enhance the surface to bulk ratio, reducing the reduction times. Moreover, a higher purity can be reached than by high pressure powder synthesis.
The IL phase LaNiO$_{2}$ can be obtained solely through topotactic reduction of the perovskite LaNiO$_{3}$ phase, which is here achieved by using CaH$_2$ as the reducing agent \cite{Wang2020,Li2020}. We reduced 50 mg of LaNiO$_3$ powder wrapped in aluminum foil, spacially separated from $250$ mg CaH$_2$ powder at 280$^\circ$C as described in detail in Ref. \cite{Ortiz2021} for various times.

We measured the stoichiometry including the hydrogen content with a combination of inductively coupled plasma mass spectroscopy (ICP-OES) and gas extraction; the former with a Spectro Ciros CCD and the latter with an Eltra ONH-2000 analyzer. For the determination of the oxygen and hydrogen content, the powder samples were placed in a Ni crucible, clipped and heated, where the  carrier  gas  takes  the  oxygen  and hydrogen out of the sample. The oxygen reacts with carbon and  CO$_{2}$  is  detected  in  a  infrared-cell, while hydrogen is detected by a thermal-conductivity-cell. Each measurement is repeated three times and compared to a standard. The quoted error bars give the statistical error.

Powder x-ray diffraction (PXRD) data were collected using a Rigaku Miniflex with a Cu K$_\alpha$ tube at room temperature.
Neutron diffraction was performed at the high-resolution powder neutron diffractometer HRPT \cite{Fischer2000} at the spallation neutron source SINQ at the Paul Scherrer Institute in Villigen. For the HRPT experiments, an amount of 340 mg of LaNiO$_{2}$ was enclosed into a vanadium can with an inner diameter of 6 mm, where the remaining space of the vanadium can was shielded with Cd foil and the measurement was carried out at 1.5 K in a $^4$He bath cryostat with a neutron wavelength of 1.15~\AA.

\section{Results}

\begin{figure}[tb]
 \begin{centering}
\includegraphics[width=1\columnwidth]{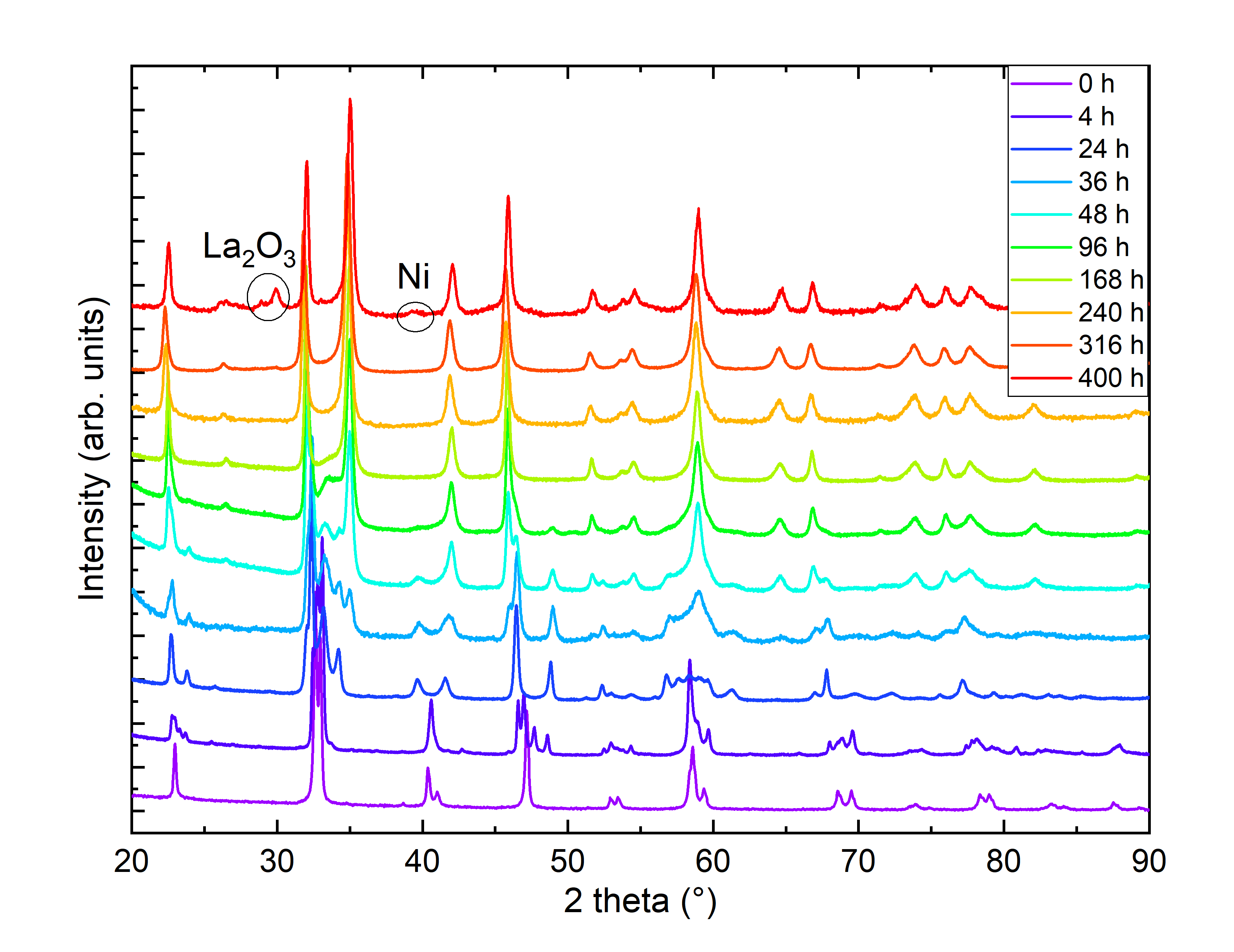}
\par\end{centering}
\caption{\textcolor{black}{\textbf{Powder x-ray diffraction (PXRD) of reduced LaNiO$_{3}$  after different reduction times.} Data were measured at $T = 300$ K with Cu K$_\alpha$ radiation. Curves are offset in vertical direction for clarity. The 0 h and 316 h data are reproduced from Ref.  \cite{Ortiz2021}.}}
\label{xrd}
\end{figure}

\begin{figure}[tb]
 \begin{centering}
\includegraphics[width=1\columnwidth]{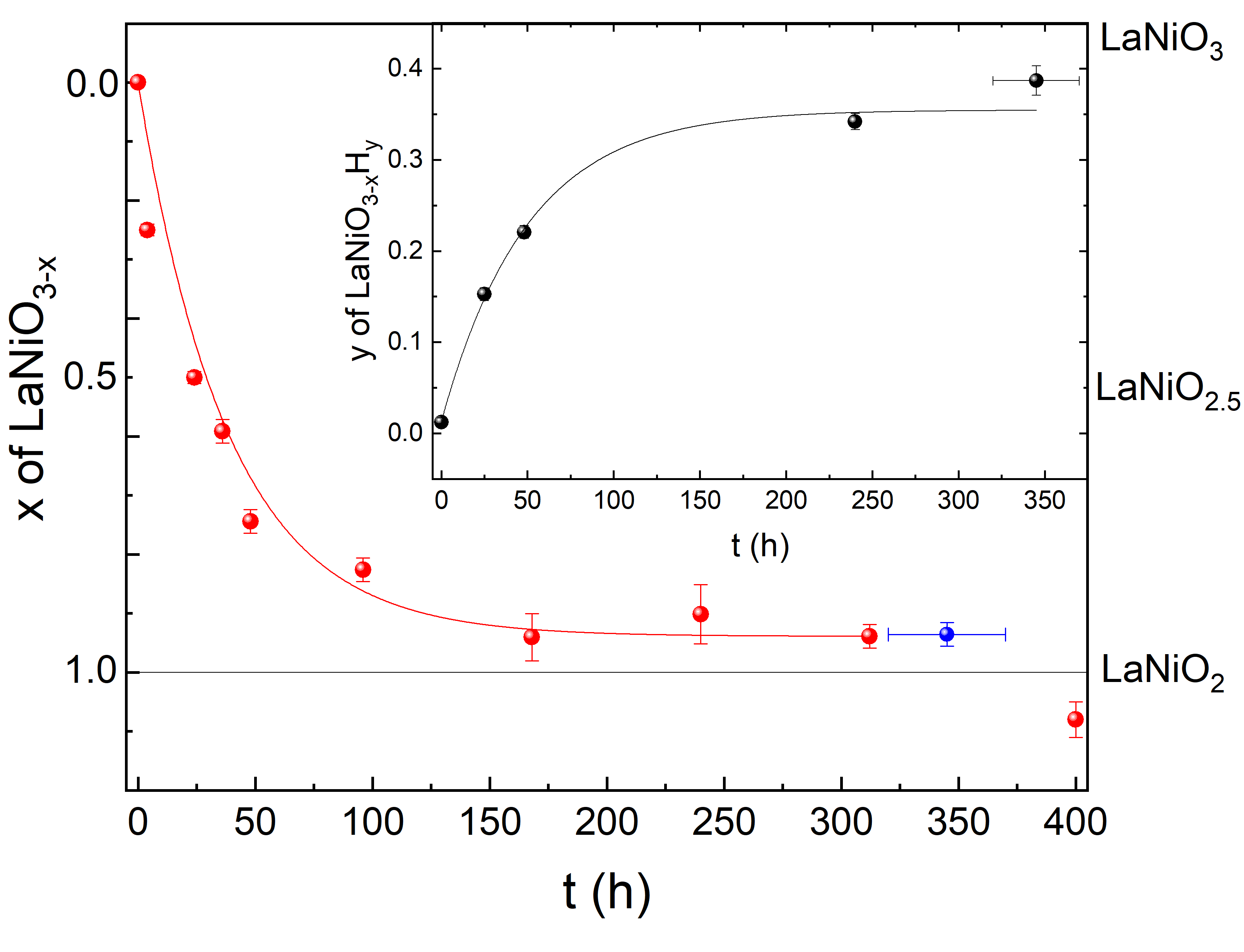}
\par\end{centering}
\caption{\textcolor{black}{\textbf{Oxygen and hydrogen content of reduced LaNiO$_{3}$ after different reduction times.} Red symbols correspond to the oxygen content $x$ in LaNiO$_{3-x}$, extracted from Rietveld refinements of the PXRD data shown in Fig.~\ref{xrd}. The blue datapoint is extracted from Rietveld refinement of our neutron diffraction data shown in Fig.~\ref{neutron}. The solid red line is a guide to the eye. The black symbols in the inset correspond to the hydrogen content $y$ in LaNiO$_{3-x}$H$_y$, obtained from gas extraction. The solid black line is a guide to the eye.}}
\label{TG}
\end{figure}

As a first step, we checked the purity and stoichiometry of our starting material in the perovskite phase. In PXRD we found no detectable impurites, and ICP combined with measurements of gas extraction indicated a starting stoichiometry of La$_{0.99(1)}$Ni$_{0.99(1)}$O$_{2.99(3)}$H$_{0.005(2)}$.
We studied the reduction progress with time in detail by preparation of several ampules and extracting samples after varying reduction times and analyzed all the products with PXRD. We noticed that the progress of the reduction depends on the purity and sample amount and grain size for a given time, as the reduction process is surface dependent. Motivated by these observations, we carried out a standardized reduction on the same batch of perovskite precursors. After one day of reduction in the described process (see Methods) we found a full structural transition to LaNiO$_{2.5}$ (see \textbf{Fig.~\ref{xrd}}). Our Rietveld refinement of the LaNiO$_{2.5}$ crystal structure is in good agreement with previous reports on reduced powder \cite{Crespin1983,Alonso1995,Alonso1997}. Notably, reduction times of one day and longer are in stark contrast to thin film samples, which can be fully reduced to the IL phase within hours \cite{Li2019,Zeng2020,Li2021G,Gao2021C,Osada2020,osada2021,zeng2021,Ikeda2016,Hepting2020}, presumably due to the enhanced thickness of our grains, with sizes of 0.5 $\mu$m as extracted from scanning electron microscopy (SEM) images. The reduction appears to be happening in domains, as we found cluster spin glass behavior in a detailed magnetic characterization \cite{Ortiz2021} and thus after longer reudction times than one day, we observe phase mixtures of LaNiO$_{2.5}$ in $P2_1/n$ (\#14) \cite{Alonso1995}, and LaNiO$_{2}$ in $P4/mmm$ (\#123) \cite{Hayward1999}, with a slowly increasing fraction of the LaNiO$_{2}$ phase as a function of time (see \textbf{Fig.~\ref{TG}}). Extracting the weight percentages from our refinement, we present the reduction progress for the CaH$_2$ reduction in analogy to a thermogravimetry (TG) curve, which is shown in Fig.~\ref{TG}. The reduction process follows an exponential decay, which we also observed in thermogravimetry studies in hydrogen gas flow (not shown here). After a reduction time of approximately 316 h, the crystal structure can be refined assuming a single phase of IL LaNiO$_{2}$ (Fig.~\ref{TG}, and Ref.~\cite{Ortiz2021}). Notably, while the structural transition does not progress further on an exponential time scale, a small amount of apical oxygen remains in the crystal lattice, as will be revealed by neutron diffraction below. However, for even longer reduction times, the sample begins to decompose, forming Ni and La$_{2}$O$_{3}$, which becomes clearly visible in PXRD after 400 h (Fig.~\ref{xrd} and Fig.~\ref{TG}).

In the inset of Fig.~\ref{TG} we plot the hydrogen content in the resulting powder samples versus the reduction time, obtained by a gas extraction method. While the initial perovskite nickelate shows a hydrogen content of 0.005(2) wt\%, we find an increase of hydrogen with time (opposite to the oxygen content), with 0.065(3) wt\% for 24 h, 0.094(3) wt\% for 38 h, 0.015(5) wt\% for 240 h and finally 0.169(7) wt\% after 320-376 h. Via gas extraction we find an oxygen content of 14.1(2) wt\% after 316 h and 14.0(2) wt\% after 320-376 h of reduction, where 14 wt\% corresponding to a full reduction to the LaNiO$_{2}$ IL phase. If we assume that the amount of hydrogen would be incorporated into the average crystal structure, the corresponding effective stoichiometry of the sample investigated with neutrons would be La$_{0.99(1)}$Ni$_{0.99(1)}$O$_{2.00(2)}$H$_{0.39(2)}$, where La and Ni are determined via ICP. However, a first hint that this might not be the case comes from the presence of small amounts of elemental Ni in the PXRD of the 400 h sample (Fig.~\ref{xrd}), which is known to trap hydrogen \cite{Louthan1975}. The Rietveld refinement of the neutron diffraction data below also indicates a small amount of Ni. Furthermore, signatures of Ni precipitates (below detection threshold of our PXRD) were also detected in powders after shorter reduction times, since they give rise to ferromagnetic contributions in the magnetic signal \cite{Ortiz2021}. The amount of Ni likely increases with reduction time, which could explain the increasing capability of incorporating hydrogen.

\begin{table}[tb]
\caption{\label{refinement} Refined atomic coordinates of LaNiO$_{2}$ in tetragonal space group $P4/mmm$ with preferred orientation extracted from powder neutron diffraction data at 1.5 K [Fig.~\ref{neutron}].}
\begin{tabular}{cccccc}
\hline
\multicolumn{6}{c}{LaNiO$_{2}$ $\vert$ $a, b = 3.9550(3)$ $\text{\AA}$, $c = 3.3588(3)$ $\text{\AA}$}\\ 
\hline
Atom & $x$ & $y$ & $z$ & $U [\text{\AA}^2]$ & Occ. \\ 
\hline
La (1d) & 0.5 & 0.5 & 0.5 & 0.0036(6) & 1 \\
Ni (1a) & 0 & 0 & 0 & 0.0057(5) & 1 \\
O (2f) & 0 & 0.5 & 0 & 0.0098(5) & 1 \\
O (1b) & 0 & 0 & 0.5 & 0.0098(5) & 0.06(1) \\
\multicolumn{5}{c}{}   \\
Reliability factors & $\chi^2$ & $R_B$ & $R_f$ & \\ \hline
 & 1.89 & 3.53 &  2.79 & \\ 
\end{tabular}
\label{table}
\end{table}

\begin{figure}[tb]
\includegraphics[width=1\columnwidth]{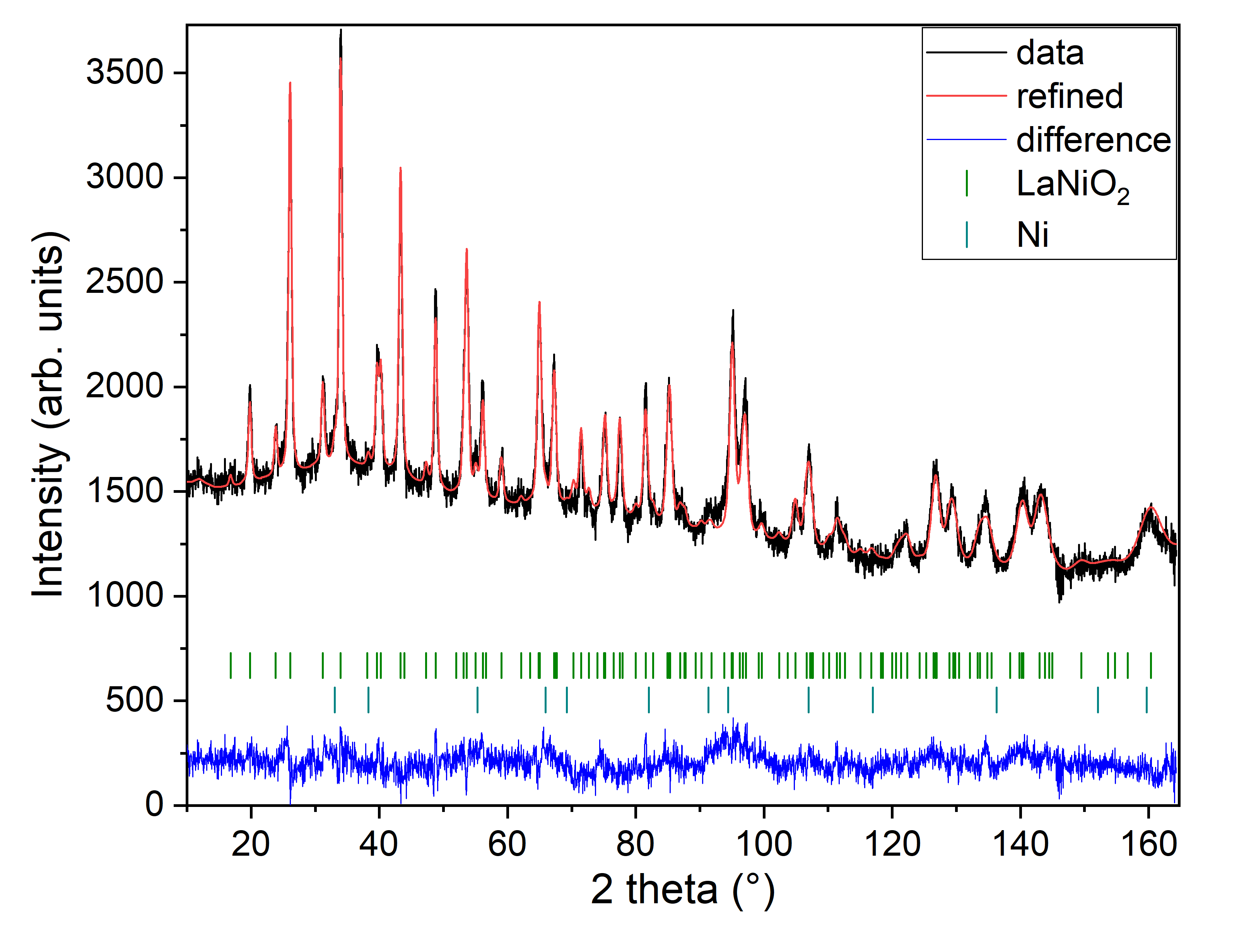}
\caption{\textcolor{black}{\textbf{Neutron powder diffraction.} The data is taken from mixed samples reduced in the range of 320-376 h. The Rietfeld refinement includes the LaNiO$_{2}$ phase and a Ni minority phase. Data were collected at 1.5 K with a neutron wavelength of 1.15~\AA.}}
\label{neutron}
\end{figure}

As a next step, we performed high resolution neutron diffraction experiments on 340 mg of a sample prepared by mixing batches that had been subject to reduction times of 320, 350 and 376 h, respectively, which are all in the range just before the start of the clear decay shown in Fig.~\ref{TG}. The obtained neutron diffraction pattern can be well refined (see \textbf{Fig.~\ref{neutron}}) assuming LaNiO$_{2}$ in the IL $P4/mmm$ structure (see Tab.~\ref{table}), with a $c$-axis parameter of 3.3588(3)~\AA, which is lower than previously reported values \cite{Ortiz2021,Hayward1999,Puphal2021}, and 1.1(2) wt\% of elemental Ni was included as a secondary phase.

\begin{figure}[tb]
\includegraphics[width=1\columnwidth]{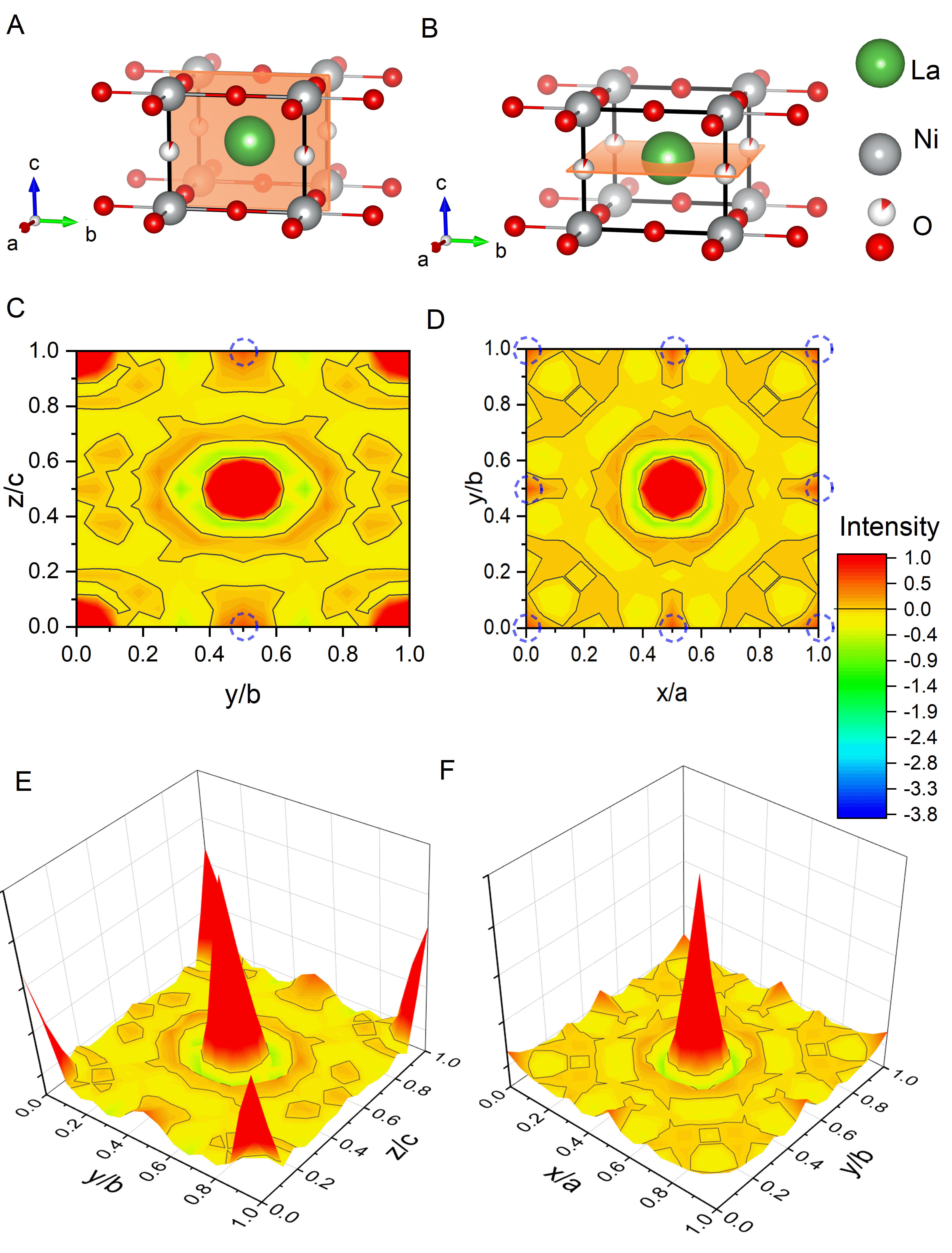}
\caption{\textbf{Fourier maps of the scattering-length density within the unit cell of LaNiO$_{2}$ extracted from neutron powder diffraction}. \textbf{A, B} Refined unit cells with cuts corresponding to the Fourier maps in \textbf{C, D}, with the intensity range focused on the negative scattering-length (yellow to blue), with -3.74~fm corresponding to hydrogen \cite{Sears1992}. The positive scattering (orange to red) is capped at 1 (with 5.8~fm corresponding to oxygen). The black lines mark the isosurface lines with zero scattering intensity. The theoretically proposed hydrogen positions \cite{Si2020,Malyi2021} are highlighted with blue dashed circles. \textbf{E, F} Fourier maps shown as a 3D colormap.}
\label{fourier}
\end{figure}

\textbf{Figure.~\ref{fourier}} shows a Fourier map of the scattering density resulting from the Rietveld refinement of the neutron data, where the range of the color scheme is chosen to focus on the negative scattering lenght of hydrogen. While the employed reduction temperatures (see Methods) are too low to enable significant mobility of the La and Ni ions, the purpose of our topotactic treatment is to reduce oxygen, which becomes mobile and is extracted by reaction with hydrogen. We therefore focus on the scattering lengths of the two elements O and H. According to the theoretical prediction of Ref.~\cite{Si2020}, intercalation of H in IL nickelates is most favorable on the vacant apical oxygen sites with Wyckoff position (1b), or on interstitial positions between the rare-earth ions (Wyckoff positions (2e) and (1c)). Another theoretical work \cite{Malyi2021} predicts that NdNiO$_{2}$ is unstable while NdNiO$_{2}$H is not, and also proposes the apical oxygen site with Wyckoff position (1b) for hydrogen. However, as shown in the Fourier maps of Fig.~\ref{fourier}, we find no significant negative scattering, which would appear as a blue color in our plots. The slightly negative scattering realized as a ring around a reflex, here in close proximity to La is an artifact from the Fourier transformation, typically seen around larger ions \cite{SanoFurukawa2018}, due to a signal cutoff from a finite q range. This is best seen in the Fourier maps Fig.~\ref{fourier} E.  
We note, that the data can be refined with similar reliability factors as in Tab.~\ref{table} by putting a mixture of O and H on Wyckoff position (1b) and constraining their occupation. As the positive scattering observed at this site could be a mixture of negative scattering H and positive scattering O, which effectively clouds the negative scattering. Realized in a substitution series with LaNiO$_{3-x}$H$_{x}$, which would yield occupations of O: 0.430(8) and H: 0.570(8) in refinements of our data, but this oxygen content is way higher than what can be found in multiple gas extraction experiments (La$_{0.99(1)}$Ni$_{0.99(1)}$O$_{2.00(2)}$H$_{0.39(2)}$). Most importantly the scenario of LaNiO$_{3-x}$H$_x$ would hint towards the possibility to synthesize the phase directly, similar as observed in the iron pnictides. However, such synthesis attempts have been carried out as described below and were not successful. Furthermore, in the case of iron pnictides the hydrogen substitution leads to no structural transitions \cite{Hanna2011,Hiraishi2014,Matsuishi2012} contrasting the evolution of the underlying structural transitions in LaNiO$_{3-x}$ according to the PXRD data (see Fig.\ref{xrd}) and literature \cite{Hayward1999}.
Another possibility to refine mixed occupancy is by constraining an equal occupation of O and H as LaNiO$_{2}$(OH)$_{x}$, which converges to 0.18(4). However, this would rather suggest an OH molecule with typical distances around 0.84~\AA, which would realize negative scattering \cite{SanoFurukawa2018} and thus reduces the quality of the fit and shifts the occupation again down to 0.06(1). Thus, we conclude that intercalated hydrogen is unlikely in our LaNiO$_{2+\delta}$ powder sample. Nevertheless, we find the presence of a small amount of residual apical oxygen in Wyckoff position (1b), which corresponds to a deviation $\delta=0.06(1)$ from the ideal stoichiometry.

Additionally, we attempted direct synthesis approaches of LaNiO$_{3-x}$H$_x$ as e.g. LaNiO$_{2}$H via a Walker-type high pressure synthesis in NaCl crucibles. We mixed La$_{2}$O$_{3}$ and NiO with NaH or CaH$_{2}$ and heated it to $650^{\circ}$C - $1000^{\circ}$C under a pressure of 5-7 GPa, but were unable to synthesize any oxyhydride nickelate. Instead we obtained La(OH)$_{3}$ and Ni, which is in contrast to related cases, such as SmFeAsO$_{1-x}$H$_{x}$ \cite{Hanna2011}, BaCrO$_{2}$H \cite{Higashi2021}, BaScO$_{2}$H \cite{Goto2017}, and SrVO$_{2}$H \cite{Yamamoto2017}.

\section{Conclusion}

In summary, we have synthesized high quality LaNiO$_{2+\delta}$ powders and found some density of residual oxygen ($\delta=0.06(1)$) even at the final stage of the topotactic reaction, just prior to decomposition of our samples. While our gas extraction method reveals partial hydrogen inclusions in the powder samples, high resolution neutron diffraction refinements show that there is no clear topotactic hydrogen in LaNiO$_{2}$, and direct attempts to synthesize LaNiO$_{3-x}$H$_x$ were unsuccessful. The hydrogen detected by gas extraction could be trapped in Ni impurities that increase with increasing reduction time and/or ascribed to phenomena at surfaces or grain boundaries such as water adsorption/intercalation \cite{Baeumer2021} on powder surfaces, or partial formation of LaNiO$_{3-x}$H$_x$ on the nm scale \cite{Onozuka2016}. Notably, a relatively high density of grain boundaries/crystallographic defects was reported for IL nickelate thin films \cite{Lee2020}, and also surface effects can play a more decisive role in films. Thus, future studies clarifying the presence and the impact of hydrogen in nickelate thin film samples are highly desirable.

\section*{Data availability statement}

The original contributions presented in the study are included in the article; further inquiries can be directed to the corresponding author.

\section*{Author Contributions}

PP, MI, BK, and MH conceived the project. RO carried out the topotactic reductions. SH executed the ICP and gas extractions. PP and VP conducted the neutron diffraction experiments. PP analyzed the data and prepared the manuscript with input from all authors.

\section*{Funding}

We acknowledge financial support by the Center for Integrated Quantum Science and Technology (IQ$^{\rm ST}$) and the Deutsche Forschungsgemeinschaft (DFG, German Research Foundation): Projektnummer 107745057 - TRR 80. The Max Planck Society is acknowledged for funding of the open access fee.
  
\section*{Acknowledgements}

We thank R. Merkle and A. Fuchs for the synthesis of LaNiO$_{3}$ and H. Hoier for preliminary PXRD characterizations. We acknowledge PXRD measurements by C. Stefani from the X-ray Diffraction Scientific Facility at an early stage of this work and S. Hammoud for carrying out the detailed ICP and gas extraction measurements. The use of
facilities of the Quantum Materials Department of H. Takagi for the attempted high pressure synthesis of LaNiO$_2$H is gratefully acknowledged.
\bibliographystyle{apsrev4-2}
\bibliography{nickelates}
\end{document}